\documentclass{edp-conf}
\usepackage{graphicx}
\begin{document}

\TitreGlobal{AGN in their Cosmic Environment}
\runningtitle{Starburst-AGN connection}
\title{Young and Intermediate Stellar population in Seyfert 2 galaxies} 
\author{Rosa M. Gonzalez Delgado}\address{Instituto de Astrofisica de Andalucia (CSIC), 
Granada, Spain}
\author{Timothy Heckman}\address{Department of Physics and Astronomy, JHU, Baltimore, USA}
\author{Claus Leitherer}\address{Space Telescope Science Institute, Baltimore, USA}
\maketitle
\begin{abstract} 
We present and discuss near-ultraviolet and optical ground-based spectra of the 20 
brightest Seyfert 2 nuclei, and HST ultraviolet images and ultraviolet spectroscopy 
for a few of them. The goal is to study the starburst-AGN connection 
and the origin of the featureless continuum. The results indicate that half
of the nuclei in the sample harbor a young and/or intermediate age population.
These stars are formed in powerful and dusty starbursts of short duration, that have 
bolometric luminosities similar to the estimated bolometric luminosities of 
their obscured Seyfert 1 nuclei. 
\end{abstract}
%
\section{Introduction}

According with the standard unified scheme for radio-quiet AGN, 
Seyfert 2 (S2) galaxies contain a Seyfert 1 (S1) nucleus that is obscured by 
a circum-nuclear torus of dust and gas; but it can be seen in polarized light. Generally 
speaking the optical continuum of the S2 nuclei can be classified in two types:
Those that show a red continuum which is very similar to that of elliptical 
galaxies, and those that show a very blue continuum in which the stellar lines 
due to an underlying bulge stellar population are very diluted. In these cases, the blue continuum 
does not show almost any stellar features and it is call features-less continuum (FC).
According with the unified picture, the blue continuum is scatered light from the 
hidden S1 nucleus. It dominates the ultraviolet continuum and contributes
between 10$\%$ and 30$\%$ to the optical light. 

However, this interpretation is not completely correct because after the 
contribution of the old stars is removed, the remaining optical continuum has a 
significant lower fractional polarization than the broad optical emission lines 
(Antonucci 1993). 
Therefore other unpolarized component has to contribute to dilute the FC.
Tran (1995) suggests that the unpolarized component is optically thin thermal 
emission from warm gas that is heated by the central hidden source. In contrast, 
Cid-Fernandes \& Terlevich (1995) and Heckman et al. (1995) suggest that the 
unpolarized component is a young very red stellar population associated with 
the obscuring torus. 

To go deeper in the origin of the FC in S2 galaxies and in the possible connection between 
Seyfert and the starburst phenomenon we have started a program to study the ultraviolet 
to near-infrared spectra of a sample of the 20 brightest S2 nuclei with the goal of
detecting stellar features from massive stars. Here, we summarize the results from the
analysis of the UV continuum of four of these objects that are presented in more detail in
Heckman et al. (1997) and Gonz\'alez Delgado  et al. (1998), and the results
from our near-ultraviolet ground based observations of the complete sample. 

\section{Sample and Observations}

We have taken ground-based optical and near infrared spectroscopy 
with the 4m telescope at Kitt Peak National Observatory for 20 of the brightest 
Seyfert 2 nuclei selected from the compilation of Whittle (1992). 
 They have been selected by their nuclear [OIII] $\lambda$5007 emission line flux 
and by their non-thermal radio continuum emission at 1.4 GHz. They all satisfy
at least one of the two following criteria: log
F$\rm_{[OIII]}\geq$ -12.0 (erg cm$^{-2}$ s$^{-1}$) and 
log F$_{1.4}\geq$-15.0 (erg cm$^{-2}$ s$^{-1}$). The criteria used guarantees 
that the whole sample is unbiased with respect to the presence or absence of a nuclear
starburst. HST UV images have been obtained for 9 of the 20 Seyfert 2
nuclei using the Faint Object Camera (FOC). We have also obtained UV spectra of 4 (Mrk 477, NGC 7130, 
NGC 5135 and IC 3639) of these 9 Seyfert 2 nuclei with the HST Goddard
High Resolution Spectrometer (GHRS). They were chosen from the subsample of 
9 Seyfert 2 nuclei for having the highest UV flux on arcsec scales.

\section{UV imaging}

Massive stars emit most of their flux at 
the UV and far-UV wavelengths. In contrast, older stars contribute significantly
to optical and longer wavelengths. Therefore, UV images provide predominantly
direct evidence of location of most of the recent unobscured star forming regions.
On the other hand, UV images of starburst galaxies show that a significant fraction
of the massive stars are formed in very compact stellar clusters (Meurer et al. 1995) that
have sizes of a few pcs. Therefore, the morphology of the UV light of S2 can 
indicate whether these nuclei harbor a starburst.

We have observed 9 objects with the FOC through the filter
F210M. This filter does not include any strong emission lines; therefore, the emission 
is dominated by the UV continuum at 2150 \AA. In four of the targets (Mrk 477, NGC 7130,
NGC 5135 and IC 3639) the morphology suggests that they harbor a nuclear starburst (Figure
1). In these cases, the UV continuum is spatially resolved, showing sub-arcsec structure.
The effective radius (50-200 pc) of the UV emission is similar to the size of
the NLR. At the position where HST+WFPC2 (F606W) suggests that the nucleus is located,
the UV images show very little emission (less than 15$\%$ of the total). This result
suggests that even at UV wavelengths the S1 nucleus is still obscured.
In other four targets, the emission is very weak and the morpholgy is very uncertain.
In contrast, in Mrk 463E, the UV emission is extended 1 Kpc and it looks
like scattered light of the hidden S1 nucleus (Figure 2).
 
\begin{figure}
\vspace{-2.5cm}
\includegraphics[width=130mm,angle=270]{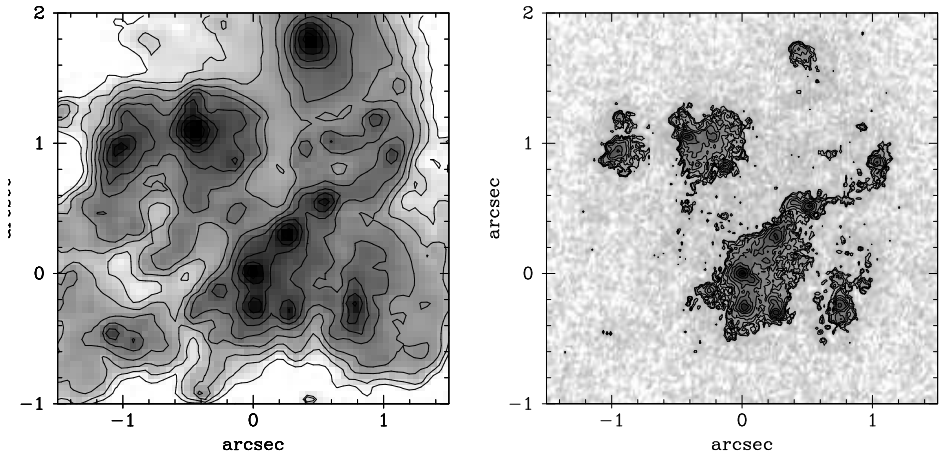}
\vspace{-4.5cm}
\caption{Central 3$\times$3 arcsec$^2$ (540 by 540 pc)
field of NGC 5135, (left): WFPC2 through F606W, (right): FOC through F210M. The origin of this plot is the pixel with
the highest UV surface brigtness. Based on the morphology of the WFPC2 image,
we believe that the nucleus is the knot located at 1.05 arcsec North and 0.4 arcsec East. 
The UV emission is resolved in knots suggesting that the nucleus of NGC 5135 harbors a starburst.}
\label{HSTuv} 
\end{figure}

\begin{figure}
\vspace{-2cm}
\center{\includegraphics[width=130mm,angle=270]{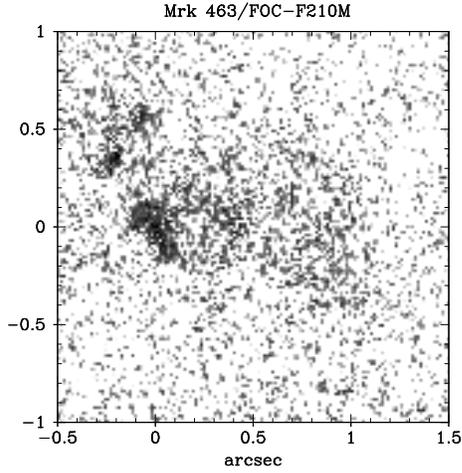}}
\vspace{-5cm}
\caption{{\it HST\/}+FOC (through F210M) image of the central 2$\times$2
arcsec$^2$ (2 by 2 Kpc) of Mrk 463E. The origin  of this plot is the brightest
UV knot which nature is unknown. Most of the UV flux comes from 1
Kpc extended emission.}  
\label{HSTuv} 
\end{figure}

\section{UV spectroscopy}

The UV spectrum of a starburst is very rich in absorption lines, many of them are formed in the 
wind of massive stars (Weedman et al. 1981). A conclusive evidence that the UV emission in the 
S2 nuclei comes from a starburst
is obtained if stellar wind resonance lines (as NV $\lambda$1240, SiIV $\lambda$1400 and 
CIV $\lambda$1540) are detected in their spectra. 
The four brighest UV S2 of our sample were observed with the GHRS and the G140L grating,
which has a nominal dispersion of 0.57 \AA/diode, using the Large Science
Aperture (LSA, 1.74$\times$1.74 arcsec) and covering 1150-1600 \AA. The spectra show
the typical absorption features of starburst galaxies, indicating 
that the UV light is dominated by the starburst component (Figure 3). Comparing the strength of the wind lines 
with those predicted by evolutionary synthesis models (Leitherer et al. 1995), we conclude that
in these S2 nuclei, the 
starbursts are of short duration with ages between 4 and 6 Myr. We find that, after 
correcting the observed UV continuum for dust extinction using the starburst attenuation law
(Calzetti et al. 1994), the bolometric luminosity of these starbursts ($\sim$ 10$^{10}$
L$\odot$) is similar to the estimated luminosity of the hidden S1 nucleus.

\begin{figure}
\vspace{3mm}
\includegraphics[width=52mm]{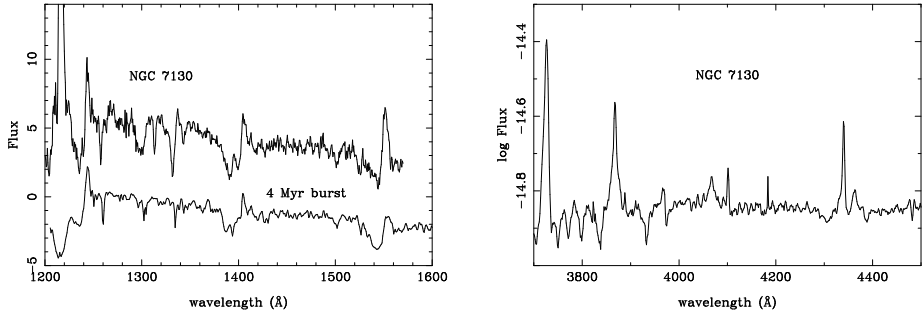}
\caption[fig]{(Left): UV (HST+GHRS) spectrum of NGC 7130 dereddened and the synthetic 4 Myr burst model (in relative units) that
fits the wind resonance stellar lines. 
The IMF slope assumed in the model is Salpeter and M$_{upp} $=80 M$\odot$. 
(Right): Ground-based optical spectrum of NGC 7130 obtained through a 1.5$\times$3.5 arcsec$^2$ aperture. It shows the higher
order terms of the Balmer series and HeI lines in absorption.}
 \end{figure}

\section{Near-ultraviolet and optical spectroscopy}

These results obtained for the brightest UV S2 of our sample indicate that these nuclei harbor
a starburst. However, they do not demonstrate that starbursts are an ubiquitous part of
the Seyfert phenomenon. To investigate deeper about the possible role that massive stars and 
starburst play in the energetics of Seyfert nuclei, we have looked for starburst signatures in the 
near-ultraviolet spectra of our whole sample. 
Massive stars also show photospheric lines at near-ultraviolet and optical wavelengths. The most notably are 
those of the H Balmer serie and HeI lines. In the spectra of starbursts, these lines are coincident 
with the nebular emission lines. However, the higher order terms of the Balmer series can be easily
detected in absorption because the strength of the Balmer series in emission decreases rapidly with 
decreasing wavelength, but the equivalent widths of the absorption lines are constant with wavelength 
(Gonz\'alez Delgado et al. 1999). Then, the detection of these absorption lines in the spectra of S2 nuclei
can indicate that these objects harbor a starburst. 

Three (NGC 5135, NGC 7130 and IC 3639) of our four S2 that 
show stellar wind resonance lines at the UV, also show the higher order terms of the Balmer series and HeI 
($\lambda$3819,4387 and 4922) in absorption. In the other object, Mrk 477, the
Balmer and HeI stellar features are overwhelmed by the corresponding nebular
lines. However, it shows a very broad HeII $\lambda$4686 emission line that is
formed in the stellar wind of Wolf-Rayet stars. These results indicate that a
method to infer the presence of a starburst in the nucleus of S2 is through the
detection of the higher order terms of the Balmer series and  HeI lines in
absorption and the Wolf-Rayet feature at 4680 \AA.
The inspection of our ground-based spectra show that six of our 20 targets show these photospheric 
features in their nuclear spectra (e.g. Mrk 1066, see Figure 4). In three more objects the stellar 
features are detected when the
contribution of the nebular emission lines are subtracted from the spectra. Thus, we conclude
that at least about half of the S2 of our sample show starburst characteristics in their
near-ultraviolet spectra. In the other half of the sample, the optical continuum show only
stellar features due to an underlying old bulge stellar population.

In order to investigate more about the origin and the contribution of the FC to the near-ultraviolet
optical continuum, we have measured the variation of the equivalent widths (Ew) of the absorption
line CaII $\lambda$3933 and G band at 4300 \AA\ toward the nucleus. In
the objects where we detect the higher order terms of the Balmer series in
absorption, the Ew of the CaII and G band decreases
significantly toward the nucleus (e.g. in NGC 5135, Figure 5). In the other
objects, they do not show dilution at all or it is very small (less than
10$\%$) (e.g. in Mrk 3, Figure 5). These results suggest that the origin of the
FC in half of the sample is produced by a nuclear starburst. In the other half,
the FC is weak relative to the light from a normal old bulge stellar population
and their origin is not clear, but it could be produced by scattered light
from the hidden S1 nucleus, because their contribution to the optical light is
similar to the fractional polarization degree detected in these types of
objects. 

\begin{figure}
\includegraphics[width=125mm]{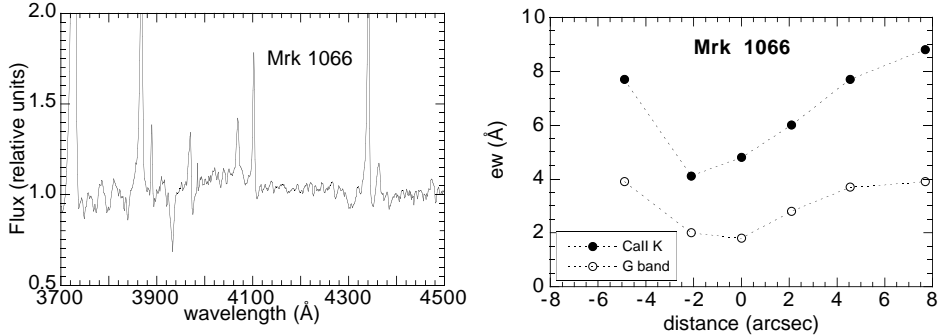}
\caption{(Right): Ground-based optical spectrum of Mrk 1066. It shows the higher
order terms of the Balmer series in absorption, as NGC 7130. (Left): Spatial distribution of the equivalent widths
of the stellar lines CaII K and G band of Mrk 1066. Note the dilution
of these lines toward its nucleus (origin of the plot).} 
\label{mrk1066} 
\end{figure}

\begin{figure}
\includegraphics[width=125mm]{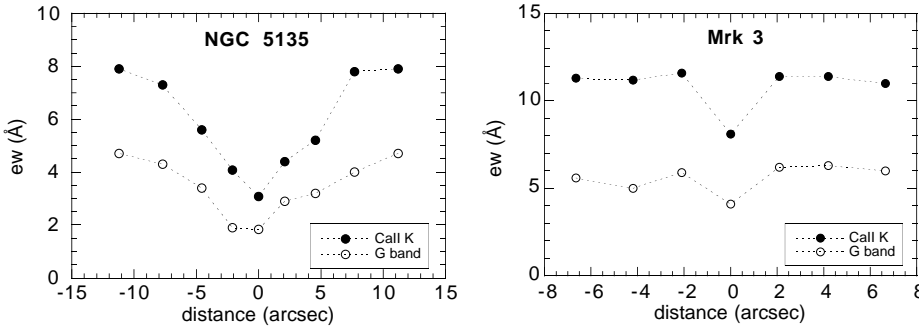}
\vspace{2mm}
\caption{As Figure 4 for NGC 5135 (left) and Mrk 3(right). 
The strong dilution of the stellar lines toward the nucleus in NGC 5135 is due to the existence of 
a starburst, but in Mrk 3 it is probably due to scattered light from the hidden S1 nucleus.  } 
\label{Ew} 
\end{figure}

Two exceptions to this dual clasification are Mrk 477 and Mrk 463E. Both objects show a very blue continuum,
very diluted stellar absorption features and very broad HeII $\lambda$4686 emission (Figure 6). In the
case of  Mrk 477, our UV images and spectroscopy indicate that it harbors a powerful red nuclear
starburst, and the HeII emission is due to Wolf-Rayet stars. In the case of Mrk 463E, the
interpretation is more controversial.  
Like in Mrk 477, in Mrk 463E, the polarized spectrum does not show HeII emission, and the fractional
degree of polarization is only a few percent. Then, the origin of the broad emission at 
$\lambda$4686 can not be due to scaterred light from the hidden S1 nucleus; therefore, it could 
due to Wolf-Rayet stars, as in Mrk 477. However, the morphology of the UV emission of Mrk 463E is
very different to that of those S2 that show starburst characteristics (Figure 1). 
Thus, we have to conclude that the origin of its FC is an enigma and Mrk 463E is a very
challanging object. HST+STIS observations will help to futher inquire about the origin of the
FC in this object.

\begin{figure}
\includegraphics[width=120mm]{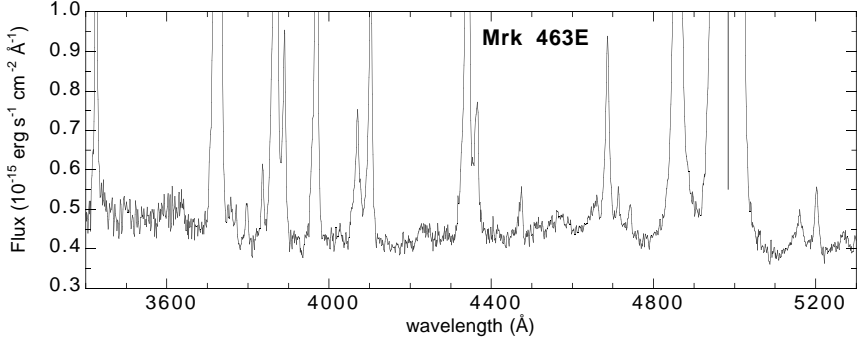}
\vspace{-12.5cm}
\caption{Ground-based optical spectrum of Mrk 463E. It shows a broad emission feature at 
4680 \AA\ which origin is unkown but that it could be produced by Wolf-Rayet stars.} 
\label{Mrk463} 
\end{figure}

\section{Conclusions}

Our near-ultraviolet spectroscopy indicate that at least half of the brighest S2
show the higher order terms of the Balmer series and HeI lines in absorption.
The analysis of the profiles of these lines using evolutionary synthesis
models  indicate that at near-ultraviolet wavelengths, the continuum is
dominated by young and intermediate age population (Figure 7). The UV images
and spectroscopy suggest that these Seyfert 2 galaxies harbor a nuclear
starburst that is responsible for the FC in these objects. These nuclear
starbursts are red and powerful. Their bolometric luminosities are similar
to the estimated luminosities of their hidden S1 nuclei. This suggests that
more powerful AGN may be related to more powerful  central starburst. In the
other half of the sample, the FC is weak and their origin may be related  to
scattered light from their obscured S1 nuclei or to nuclear starbursts that is
 much less conspicuous. These results are also in agreement with those found
by Cid-Fernandes et al. (1998) and Schmitt et al. (1999) in a related work
analyzing  S1, S2 and Liners. They found that in 40$\%$ of the S2 targets of
their sample the nuclear spectrum show stellar signatures from
a young and/or intermediate age population.

\begin{figure}
\vspace{-0.5cm}
\includegraphics[width=100mm]{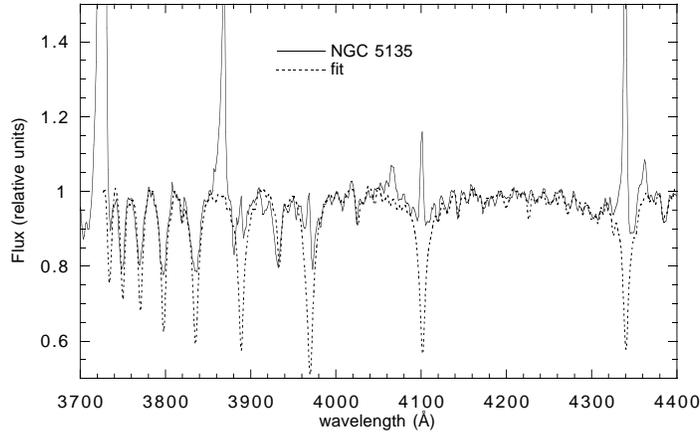}
\caption{Ground-based optical spectrum of NGC 5135 and the synthetic composite model that fit the 
profile of the stellar Balmer and HeI lines, and the strength of the CaII K and G band. 
The model has the contributions of a young starburst (3 Myr old), and an
intermediate (a few hundred Myr old) and an old stellar population. The
relative contributions of these components to the total light at 4400 \AA\ is
50$\%$, 40$\%$ and 10$\%$, respectively.}  \label{fit5135} 
\end{figure}

Two special cases are Mrk 477 and Mrk 463E. They both show a very blue continuum,
the stellar features very diluted and broad emission at 4680 \AA. Recently, 
this broad emission has been found in other Seyfert galaxies (Storchi-Bergmann 
et al. 1998, Tran et al. 1999, Contini these proceedings).
In all these cases, the broad emission is attributed to Wolf-Rayet stars. However,
like in Mrk 463E, there is not an additional evidence of the presence of a starburst
in the nucleus of these Seyferts. The exception is Mrk 477 (Heckman et al. 1997) where
we detected ultraviolet resonance lines formed in the wind of massive stars. Additional
HST+STIS observations are necessary to confirm that Mrk 463E and similar objects are 
really Wolf-Rayet-Seyfert galaxies.


\end{document}